\documentclass{article}
\usepackage[utf8]{inputenc}
\pdfoutput=1
\usepackage{amsmath}
\usepackage{enumerate}
\usepackage{amssymb}
\usepackage{float}
\usepackage{graphicx}
\usepackage{multirow}
\usepackage[usenames,dvipsnames]{xcolor}
\usepackage{bm}
\usepackage[normalem]{ulem}
\usepackage[autostyle]{csquotes} 
\usepackage{tikz}
\usetikzlibrary{positioning,shapes.geometric, arrows, arrows.meta}

\usepackage{booktabs}
\usepackage{color}
\usepackage{jcappub}

\def\vk{{\bf k}}
\def\fnl{f_{\rm NL}}

\title{\texttt{GENGARS}: Accurate non-Gaussian initial conditions with arbitrary bispectrum for N-body simulations}
\author[1]{Emanuele Fondi,}
\author[1,2]{Licia Verde,}
\author[3,4,5]{Marco Baldi,}
\author[6,7]{William Coulton,}
\author[8,9]{Francisco Villaescusa-Navarro,}
\author[10,11]{Benjamin Dan Wandelt}

\affiliation[1]{Institut de Ciències del Cosmos, Universitat de Barcelona (ICCUB), Martí i Franquès, 1, 08028 Barcelona, Spain}
\affiliation[2]{ICREA, Pg. Lluis Companys 23, Barcelona, 08010, Spain} 
\affiliation[3]{Dipartimento di Fisica e Astronomia, Alma Mater Studiorum Universit\`a di Bologna, via Piero Gobetti, 93/2, I-40129 Bologna, Italy;}
\affiliation[4]{Osservatorio di Astrofisica e Scienza dello Spazio, via Piero Gobetti 93/3 1, I-40129 Bologna, Italy;}
\affiliation[5]{INFN - Sezione di Bologna, viale Berti Pichat 6/2, I-40127 Bologna, Italy;}
\affiliation[6]{ Kavli Institute for Cosmology, Madingley Road, Cambridge, CB3 0HA, UK}
\affiliation[7]{DAMTP, Centre for Mathematical Sciences, Wilberforce Road, Cambridge CB3 0WA, UK}
\affiliation[8]{Center for Computational Astrophysics, 162 5th Avenue, New York, NY, 10010, USA}
\affiliation[9]{Department of Astrophysical Sciences, Princeton University, 4 Ivy Lane, Princeton, 
NJ 08544 USA}
\affiliation[10]{Department of Physics and Astronomy, Johns Hopkins University, Baltimore, MD 21218}
\affiliation[11]{Department of Applied Mathematics and Statistics, Johns Hopkins University, Baltimore, MD 21218}

\emailAdd{emanuelefondi@icc.ub.edu, liciaverde@icc.ub.edu}

\abstract{Primordial non-Gaussianity is predicted by various inflationary models, and N-body simulations are a crucial tool for studying its imprints on large-scale structure. In this work, we present \texttt{GENGARS} ( GEnerator of Non-Gaussian ARbitrary Shapes), a framework for generating accurate non-Gaussian initial conditions for N-body simulations. 
It builds upon the formulation introduced by Wagner \& Verde (2012), enabling to generate a primordial gravitational potential with a desired separable bispectrum $B_{\Phi}(k_1,k_2,k_3)$. For the local, equilateral and orthogonal non-Gaussian templates, we benchmark our method against the well-established \texttt{2LPT-PNG} code. We show that \texttt{GENGARS} achieves improved accuracy and lower noise by suppressing spurious contributions to the primordial power spectrum. This paper aims at presenting the method, quantifying its performance and illustrating the benefits and applicable use cases over existing approaches.
}

\begin{document}
\maketitle
\flushbottom

\section{Introduction}
The inflationary paradigm represents the leading theoretical framework for explaining the origin of primordial fluctuations that seeded the formation of large-scale structures in the Universe. In its simplest form—single-field slow-roll inflation with canonical kinetic energy and initial Bunch-Davies vacuum—the resulting primordial fluctuations are predicted to be nearly Gaussian, with deviations from Gaussianity expected to be exceedingly small \citep{maldacena_non-gaussian_2003, acquaviva_second-order_2002, creminelli_single_2004}. However, relaxing any of these standard assumptions naturally leads to significant levels of primordial non-Gaussianity (PNG), whose statistical properties encode critical information about the physics of the early Universe (see e.g. \cite{bartolo2004, chen2010, Meerburg:2019qqi,Chen2007,Chen2010_2,Chen2010_3,Arkani-Hamed_2004,Alishahiha:2004eh,Langlois2008,LindeMukhanov1997,Clifford2008}, and references therein). At leading order, different inflationary mechanisms leave distinct imprints in the primordial three-point correlation function or bispectrum, characterized by their shape and parameterized by the amplitude $f_{\rm NL}$. While higher-order correlators such as the trispectrum also contain valuable information, in this work we focus on the bispectrum, which we assume fully specifies the PNG template. Observational constraints on these shapes provide powerful tests of fundamental physics at energies far beyond the reach of terrestrial experiments \citep{Babich:2004gb, ArkaniHamedMaldacena2015, Achucarro:2022qrl}.

While the Cosmic Microwave Background (CMB) has provided stringent constraints on PNG \cite{Planck:2019kim}, ongoing and forthcoming galaxy surveys hold the potential to greatly improve these constraints \cite[e.g.,][]{Alvarez14, desjacques10, Sailer:2021yzm, Shiveshwarkar:2023afl, chaussidon_fnl, Euclid2020pxf}. Realizing this potential, however, requires accurate theoretical predictions that capture the complex non-linear gravitational evolution of matter fluctuations \cite{karagiannis_constraining_2018, Cabass:2022wjy, DAmico:2022gki, Cabass:2024wob}. In this context, cosmological N-body simulations have thus emerged as an indispensable tool for interpreting observational data, as they fully capture the non-linear dynamics of structure formation. They provide a controlled framework to investigate the imprints and detectability of different PNG shapes, allowing forecasts and simulation-based analyses of the large-scale structure \cite{Coulton:2022qbc, Coulton:2022rir, Andrews:2022nvv,Jung2023, Jung24, Baldi24, Goldstein25, Stahl:2024stz,Adame:2023nsx}.

N-body simulations trace the gravitational dynamics of discrete dark matter particles from their initial distribution, defined in a comoving cubic box. Particle positions and velocities are typically initalized using Lagrangian Perturbation Theory (LPT) \cite{Peebles_1980,bernardeau2002,Sirko2005,crocce2006}, which requires specifying the primordial gravitational potential $\Phi$. For standard initial conditions, this is drawn from a Gaussian distribution, while for PNG simulations, $\Phi$ must be constructed to reflect the desired primordial bispectrum. A widely used method to generate non-Gaussian initial conditions for specific templates- local, equilateral, and orthogonal - was introduced by Ref.~\cite{Scoccimarro12}. This approach employs Fast-Fourier Transforms (FFT), making it fast and numerically affordable. However, its implementation is linked to the structure of these specific templates and cannot be directly generalized to other physically-motivated bispectra \cite{Planck:2019kim, Euclid:2023shr, Babich2004, Renaux-Petel2009}. The formulation by Wagner \& Verde \citep{Wagner:2011wx, Wagner_Verde_Boubekeur_2010} offers a more general and flexible approach by directly employing the \textit{reduced bispectrum kernel} \cite{Schmidt:2010gw}. This kernel can be constructed directly from the target bispectrum shape. Despite its generality and theoretical advantages, the implementation originally proposed by Ref.\cite{Wagner:2011wx} was highly demanding, scaling prohibitively with simulation resolution.

In this paper, we present \texttt{GENGARS} (\textbf{GE}nerator of \textbf{N}on-\textbf{G}aussian \textbf{AR}bitrary \textbf{S}hapes), a code designed to overcome these limitations by employing a mathematically equivalent—but computationally optimized—formulation of the Wagner \& Verde kernel. Our approach exploits a separable decomposition of the bispectrum kernel through the Schwinger parameterization \cite{Smith2011}, significantly reducing the computational cost. This formulation allows us to use FFTs to efficiently generate initial conditions for arbitrary separable primordial bispectra, opening new avenues for testing non-standard inflationary scenarios through N-body simulations.

To assess the accuracy and computational performance of \texttt{GENGARS}, we benchmark our implementation against the well-established \texttt{2LPT-PNG}\footnote{\url{https://github.com/dsjamieson/2LPTPNG}} code \citep{Scoccimarro12} for the local, equilateral, and orthogonal templates. We compare both the initial conditions and the evolved matter and halo statistics at $z=0$, highlighting the accuracy of our novel implementation and the differences with \texttt{2LPT-PNG}. By making \texttt{GENGARS} publicly available\footnote{\url{https://github.com/emanuelefondi/GENGARS}} upon publication, we aim to provide the community with a flexible tool for exploring the imprint of PNG on large-scale structure, ultimately assessing the detectability of primordial features through ongoing and forthcoming galaxy surveys.

The paper is structured as follows. In Section 2, we review the theoretical foundations of non-Gaussian initial conditions generation and present the formulation underlying our approach. Section 3 details the implementation of \texttt{GENGARS} and its code structure. In Section 4, we compare initial conditions generated with \texttt{GENGARS} and \texttt{2LPT-PNG}, focusing on the primordial power spectrum, bispectrum, and computational performance. Section 5 presents an application to a non-standard bispectrum shape with oscillatory features, showcasing the flexibility of our method. In Section 6, we compare the $z=0$ matter power spectrum, bispectrum, and halo mass function obtained from N-body simulations initialized with \texttt{GENGARS} and \texttt{2LPT-PNG}. We finally summarize our results in Section 7.

\section{Theoretical background}
\label{sec:theory}
In the standard scenario, inflation is driven by a single scalar field undergoing slow-roll evolution from an initial Bunch-Davies vacuum, with canonical kinetic energy. Relaxing (at least) one of these assumptions has an impact on the statistical properties of the primordial gravitational potential fluctuations $\Phi$, which inherit non-Gaussian features. The lowest-order statistic sensitive to departures from Gaussianity is the three-point function, or equivalently, its Fourier counterpart, the bispectrum $B_{\Phi}(\mathbf{k_1},\mathbf{k_2},\mathbf{k_3})$, defined by:

\begin{equation}\label{eq:bispectrumdef}
\langle \Phi(\mathbf{k}_1) \Phi(\mathbf{k}_2) \Phi(\mathbf{k}_3) \rangle = (2\pi)^3 \delta_D(\mathbf{k}_1 + \mathbf{k}_2 + \mathbf{k}_3) B_{\Phi}(k_1,k_2,k_3),
\end{equation}
where the Dirac delta function ensures statistical homogeneity, while isotropy implies that the bispectrum depends only on the magnitudes of the wavevectors. Different PNG shapes emerge from different classes of inflationary models, each associated with specific physical mechanisms \cite{bartolo2004, chen2010, Meerburg:2019qqi,Chen2007,Chen2010_2,Chen2010_3,Arkani-Hamed_2004,Alishahiha:2004eh,Langlois2008,LindeMukhanov1997,Clifford2008}. The functional form of the primordial bispectrum $B_{\Phi}(k_1,k_2,k_3)$ predicted by inflationary models can be quite complex. Thus, for computational purposes, physical shapes are often approximated by templates, constructed to maximize their correlation with the original shapes across all triangle configurations \cite{Babich2004}.\\

A well-studied template is the local one, which arises in multi-field inflationary models. This shape is characterized by a strong signal in the squeezed limit $k_1 \ll k_2 \approx k_3$ and takes the form:

\begin{equation}
\label{eq:b_loc}
B_{\Phi}^{\text{loc}}(k_1, k_2, k_3) = 2 f_{\text{NL}}^{\text{loc}} \left[P_{\Phi}(k_1) P_{\Phi}(k_2) + P_{\Phi}(k_2) P_{\Phi}(k_3) + P_{\Phi}(k_1) P_{\Phi}(k_3)\right],
\end{equation}
where $P_{\Phi}(k)=\frac{18 \pi^2}{25} A_s \frac{k^{n_s-4}}{k_p^{n_s-1}}$ is the primordial power spectrum. This template is actually exact at leading order for a \textit{local} quadratic correction to the Gaussian primordial potential $\Phi=\phi+f^{\rm loc}_{\rm NL}(\phi^2-\left<\phi^2\right>) $ with $\phi$ a Gaussian random field.
If the inflationary dynamics involve non-canonical kinetic terms, such as in the Dirac-Born-Infeld (DBI) inflation \cite{Alishahiha:2004eh}, or in ghost inflation \cite{Arkani-Hamed_2004}, the resulting primordial bispectrum is well approximated by the equilateral template \cite{Clifford2008}. This shape peaks in the equilateral configuration $k_1 \approx k_2 \approx k_3$ and is given by:

\begin{equation}
\label{eq:b_eq}
\begin{aligned}
B_{\Phi}^{\text{eq}}(k_1, k_2, k_3) = & \ 6 f_{\text{NL}}^{\text{eq}} \bigg[ -\left(P_{\Phi}(k_1) P_{\Phi}(k_2) + \text{2 perm.} \right) - 2 \left(P_{\Phi}(k_1) P_{\Phi}(k_2) P_{\Phi}(k_3)\right)^{2/3} \\
& + \left(P_{\Phi}(k_1)^{1/3} P_{\Phi}(k_2)^{2/3} P_{\Phi}(k_3) + \text{5 perm.}\right) \bigg].
\end{aligned}
\end{equation}

Inflationary models with an initial state different from the Bunch-Davies vacuum, i.e. excited initial states \cite{Holman:2007na}, generate primordial bispectra which peak for flattened ($k_1 \approx k_2 + k_3$) configurations. These are well described by a linear combination of the equilateral and the so-called orthogonal template. The latter is constructed to be \textit{orthogonal} \cite{Senatore_etal_2010} to both local and equilateral templates, and takes the form:

\begin{equation}
\label{eq:b_ort}
\begin{aligned}
B_{\Phi}^{\text{ort}}(k_1, k_2, k_3) = & \ 6 f_{\text{NL}}^{\text{ort}} \bigg[ -3 \left(P_{\Phi}(k_1) P_{\Phi}(k_2) + \text{2 perm.} \right) \\&- 8 \left(P_{\Phi}(k_1) P_{\Phi}(k_2) P_{\Phi}(k_3)\right)^{2/3} \\
& + 3\left(P_{\Phi}(k_1)^{1/3} P_{\Phi}(k_2)^{2/3} P_{\Phi}(k_3) + \text{5 perm.}\right) \bigg].
\end{aligned}
\end{equation}

Although these templates provide useful approximations to broad classes of inflationary models, they do not fully capture the diversity of possible bispectrum shapes originating from several well-motivated scenarios. For instance, models with sharp features or resonances in the inflaton potential can generate oscillatory bispectra \cite{chen2010, Chen:2010bka,Euclid:2023shr}. Similarly, oscillations characterize \textit{cosmological collider} models, in which the inflaton is coupled to massive fields \cite{ArkaniHamedMaldacena2015}. Additionally, scenarios with a time-dependent sound speed or other deviations from standard slow-roll inflation may lead to running non-Gaussianity \cite{LoVerde_etal_2008,Sefusatti_etal_2009}, where $f_{\rm NL}=f_{\rm NL}(k)$. Understanding the impact of these non-standard shapes in the large-scale structure, and assessing their detectability, motivates the need for a more flexible framework for generating initial conditions for simulations with PNG.

\subsection{Generating initial conditions for arbitrary bispectrum shapes}

As mentioned above, in the Gaussian case, the primordial potential is simply represented by a Gaussian random field $\Phi^{\rm G}$. Starting from a Gaussian primordial potential, PNG introduces higher-order correlations that can be expressed, at leading order, as a quadratic correction to $\Phi^{\rm G}$. The most general expression of such correction in Fourier space is given by \cite{Schmidt:2010gw}:
\begin{equation}\label{eq:generic_phi}
    \Phi(\vk)\equiv\Phi^{\rm G}(\vk)+\fnl \Phi^{\rm NG}(\vk)=\Phi^{\rm G}(\vk)+\fnl \int \frac{d^3k'}{(2\pi)^3} W(k, k', |\mathbf{k} + \mathbf{k}'|)\Phi^{\rm G*}(\vk') \Phi^{\rm G}(\mathbf{k} + \mathbf{k}')
\end{equation} 
where the kernel $W(k_1,k_2,k_3)$ only depends on the wavevector magnitudes for statistical isotropy and $k_3=|\vk_1+\vk_2|$ ensures homogeneity. This expression can be understood as a generalization of the local PNG case, for which $W\equiv 1$ and the convolution can be computed as a product in real space. For a generic $W$, the bispectrum of Eq.~\eqref{eq:generic_phi} reads
    \begin{equation}\label{eq:kernel_eq}
B_\Phi(k_1, k_2, k_3) = 2 f_{\rm NL} \left[ W(k_1, k_2, k_3) P_\Phi(k_1) P_\Phi(k_2) + \text{2 perm.} \;\right].
\end{equation}

Now, imposing $B_\Phi(k_1, k_2, k_3)$ to be our target bispectrum, we need to invert Eq.~\eqref{eq:kernel_eq} to determine the generating kernel $W$. However, this problem does not have a unique solution, i.e. there exist several choices of $W$ that lead to the same bispectrum, at leading order.

Assuming to have a simulation box with $N_g$ grid points per side, a brute-force evaluation of the triple integral in Eq.~\eqref{eq:generic_phi} for every $\mathbf{k}$ would scale as $\mathcal{O}(N_g^{\, 6})$. As a consequence, the computational cost of the direct implementation of \eqref{eq:generic_phi} to generate initial conditions becomes prohibitive very quickly even for modest $N_g$.

Exploiting the freedom in the kernel choice, a common approach to overcome the high computational cost of this calculation is to assume the kernel to be \textit{separable}, i.e., it can be written in the form
\begin{equation}\label{eq:sep_kernel}
    W(k_1, k_2, k_3) = \sum_{i=1}^{N_i} w_1^i(k_1) w_2^i(k_2) w_3^i(k_3),
\end{equation}
Under this assumption, the non-Gaussian contribution to the primordial potential in Eq.~\eqref{eq:generic_phi} becomes
\begin{equation}
\label{eq:phi_conv}
    \Phi^{\rm NG}(\mathbf{k}) = \sum_{i=1}^{N_i} w_1^i(k)\int \frac{d^3k'}{(2\pi)^3}\, w_2^i(k')\Phi^{\rm G*}(\vk')\;w_3^i(|\mathbf{k} + \mathbf{k}'|) \Phi^{\rm G}(\mathbf{k} + \mathbf{k}')
\end{equation}
where the integral is explicitly written as a convolution between two fields. The computational advantage of this formulation lies in the fact that the convolution can be performed as a product in real space and quickly computed via FFT.

While  Eq.~\eqref{eq:generic_phi} would scale as $\mathcal{O}(N_g^{\, 6})$, 
using a separable kernel and FFTs, the scaling improves to $\mathcal{O}(N_i\,N_g^3\,{\rm log}N_g)$. This method provides a substantial computational improvement, making it feasible to generate initial conditions for large-scale simulations.

\subsection{Impact of kernel choice on the primordial power spectrum}\label{subsec:kernel_choice}

Introducing a non-Gaussian correction to the primordial gravitational potential $\Phi$ as in Eq.~\eqref{eq:generic_phi} modifies the primordial power spectrum, which can now be written as
\begin{equation}
(2\pi)^3 \delta_D(\mathbf{k} + \mathbf{k'})P_{\Phi}(k)=\langle \Phi^{\rm G}(\mathbf{k}) \Phi^{\rm G}(\mathbf{k'}) \rangle + 2\langle \Phi^{\rm G}(\mathbf{k}) \Phi^{\rm NG}(\mathbf{k'}) \rangle + \langle \Phi^{\rm NG}(\mathbf{k}) \Phi^{\rm NG}(\mathbf{k'}) \rangle.
\label{eq:pk_phi_terms}
\end{equation}
From a theoretical point of view, the mixed term $2\langle \Phi^{\rm G}(\mathbf{k}) \Phi^{\rm NG}(\mathbf{k'}) \rangle$ vanishes since it involves an odd number of Gaussian fields \cite{Wagner_Verde_Boubekeur_2010}. However, this cancellation only holds in the limit of infinite resolution or when averaging over many realizations. In practice, this term does not contribute to the mean but still increases the variance of $P_{\Phi}(k)$.
On the other hand, the contribution $\langle \Phi^{\rm NG}(\mathbf{k}) \Phi^{\rm NG}(\mathbf{k'}) \rangle = (2\pi)^3 \delta_D(\mathbf{k} + \mathbf{k'}) P_{\Phi}^{\rm NG}(k)$ is not vanishing and can be significant depending on the kernel choice \cite{Wagner:2011wx,Scoccimarro12,Wagner_Verde_Boubekeur_2010}. In general, since the primordial potential is well constrained on large scales by CMB observations, one requires the non-Gaussian correction \( P_{\Phi}^{\rm NG}(k) \) to be small at low \( k \), to avoid violating observational bounds (e.g., on $n_s$) \cite{Planck_2018_VI}. By using Eq.~\eqref{eq:generic_phi}, the non-Gaussian power spectrum correction for a generic kernel $W(k_1, k_2, k_3)$ is given by:

\begin{equation}
\label{eq:ng_pk_contribution}
P_{\Phi}^{\rm NG}(k) = 2 f_{\rm NL}^2 \int \frac{d^3 k'}{(2\pi)^3} W^2(k, k', |\mathbf{k} + \mathbf{k}'|) P^{\rm G}_{\Phi}(k') P^{\rm G}_{\Phi}(|\mathbf{k} + \mathbf{k}'|).
\end{equation}
The large-scale behavior of Eq.~\eqref{eq:ng_pk_contribution} is determined by $\lim_{k \to 0} W^2(k, k', |\mathbf{k} + \mathbf{k}'|)$, i.e., by the squeezed limit of the kernel. Imposing that the kernel does not contain contributions that alter the large-scale scaling of $P_{\Phi}(k)$ provides a criterion to reduce the freedom in the kernel choice. This strategy is adopted in \cite{Scoccimarro12}, where for each bispectrum template, a separable kernel solution to Eq.~\eqref{eq:kernel_eq} is first expressed in terms of free parameters, which are then fixed by requiring that $P_{\Phi}^{\rm NG}(k)$ does not diverge faster than $P^{\rm G}_{\Phi}(k)\sim k^{-3}$ at low $k$. Beyond the scaling, one should also pay attention to the amplitude of $P_{\Phi}^{\rm NG}(k)$, governed by the kernel dependence on $k'$. In particular, the integral can develop an infrared ($k'\rightarrow0$) divergence. In a finite simulation box, this infrared behavior is regularized by the minimal mode $k_{\min} = 2\pi/L$, but the integral can still receive large contributions around small $k'$. As a result, even kernels that respect the desired large-scale scaling in $k$ can lead to an enhanced amplitude of $P_{\Phi}^{\rm NG}(k)$ due to their structure in $k'$.

For example, for the local case with a kernel $W\equiv1$, the integral becomes a renormalization of the amplitude and $P_{\Phi}^{\rm NG}(k) \propto P_{\Phi}(k)$ \cite{mcdonald_primordial_2008}. This corresponds to a change in the slope of the primordial power spectrum. However, even for an unrealistically large value of $f_{\rm NL}=1000$, this change is $\Delta n_s \simeq 0.003$ \cite{mcdonald_primordial_2008}, which remains within Planck constraints on $n_s$ \cite{Planck_2018_VI}. On the other hand, using a local PNG kernel different from $W\equiv1$, as the one adopted in Ref.~\cite{Wagner_Verde_Boubekeur_2010}, we can have $P^2_{\Phi}(k)$ contributions with a large amplitude, which introduce large modifications to $n_s$.

Given these considerations, an optimal kernel should be separable and minimize the impact on the large-scale primordial power spectrum.

\subsection{The reduced bispectrum kernel in separable form}
The choice of Ref.~\cite{Wagner:2011wx} for the kernel $W$ consists in the reduced bispectrum of the target:
\begin{equation}\label{eq:wagner_kernel}
    W(k_1,k_2,k_3)=\frac{B_\Phi(k_1, k_2, k_3)}{2 f_{\text{NL}} \left[P_{\Phi}(k_1) P_{\Phi}(k_2) + P_{\Phi}(k_2) P_{\Phi}(k_3) + P_{\Phi}(k_1) P_{\Phi}(k_3)\right]}.
\end{equation}
This choice uniquely fixes the solution of Eq.~\eqref{eq:kernel_eq}, allowing for the generation of initial conditions for arbitrary bispectra, as the kernel is explicitly constructed from the target bispectrum $B_{\Phi}$.
Furthermore, the denominator peaks in the squeezed limit, suppressing $\lim_{k \to 0} W^2(k, k', |\mathbf{k} + \mathbf{k}'|)$ in Eq.~\eqref{eq:ng_pk_contribution} by a factor $\sim k^6$. This allows us to automatically regularize the large-scale scaling of $P_{\Phi}^{\rm NG}(k)$ without requiring additional tuning. At the same time, this kernel also regularizes the infrared behavior in $k'$, thus taking the amplitude of $P_{\Phi}^{\rm NG}(k)$ under control. We refer the reader to Appendix A of Ref.\cite{Wagner_Verde_Boubekeur_2010} for an explicit demonstration for the templates in Eqs.~\eqref{eq:b_loc}-\eqref{eq:b_ort}. 

Because of the denominator, however, this kernel is not written in a separable form as in Eq.~\eqref{eq:sep_kernel}. As a result, a direct implementation of this prescription through Eq.~\eqref{eq:generic_phi} leads to a prohibitive computational cost that scales as $\mathcal{O}(N_g^{\, 6})$, as mentioned above.

To overcome this challenge, we introduce the so-called Schwinger parameterization \cite{Smith2011}, that allows us to rewrite the denominator in Eq.~ \eqref{eq:wagner_kernel} in a separable form. This follows from the key identity
\begin{equation}
    \frac{1}{f(x)} = \int_0^{\infty} e^{-t f(x)} dt.
\end{equation}
Applying this identity to the denominator of Eq.~ \eqref{eq:wagner_kernel} we obtain 
\begin{equation}
\label{eq:denom_schwinger}
    \frac{1}{P_{\Phi}(k_1) P_{\Phi}(k_2) + P_{\Phi}(k_2) P_{\Phi}(k_3) + P_{\Phi}(k_1) P_{\Phi}(k_3)}= \int_0^{\infty} dt \;\frac{e^{-\frac{t}{P_{\Phi}(k_1)}}}{P_{\Phi}(k_1)}\frac{e^{-\frac{t}{P_{\Phi}(k_2)}}}{P_{\Phi}(k_2)}\frac{e^{-\frac{t}{P_{\Phi}(k_3)}}}{P_{\Phi}(k_3)}.
\end{equation}
In this way, the denominator is rewritten as an integral over a new variable $t$, with the integrand written in a separable form. If the target bispectrum $B_{\Phi}$ is also separable, as for the templates in Eqs.~\eqref{eq:b_loc}-\eqref{eq:b_ort}, then it can be expressed as
\begin{equation}
\label{eq:sep_bk}
    B_{\Phi}(k_1, k_2, k_3) = \sum_{i=1}^{N_i} b_1^i(k_1) b_2^i(k_2) b_3^i(k_3),
\end{equation}
with $N_i$ being the number of terms which define the template, e.g. $N_i=10$ for the equilateral template \eqref{eq:b_eq}. By combining Eqs.~\eqref{eq:wagner_kernel},\eqref{eq:denom_schwinger},\eqref{eq:sep_bk}, we can then rewrite the non-Gaussian contribution in Eq.~\eqref{eq:generic_phi} as 
\begin{equation}
\label{eq:phi_conv_schwinger}
    \Phi^{\rm NG}(\mathbf{k}) = \int_0^{\infty} dt \;\sum_{i=1}^{N_i} w_1^i(t,k)\int \frac{d^3k'}{(2\pi)^3}\, w_2^i(t,k')\Phi^{\rm G*}(\vk')\;w_3^i(t,|\mathbf{k} + \mathbf{k}'|) \Phi^{\rm G}(\mathbf{k} + \mathbf{k}')
\end{equation}
where
\begin{equation}
    w_j^i(t,k)=b_j^i(k)\, \frac{e^{-\frac{t}{P_{\Phi}(k)}}}{P_{\Phi}(k)} \hspace{2cm} \text{for j=1,2,3}.
\end{equation}
With respect to Eq.~\eqref{eq:phi_conv},  the additional overhead is related to the integral over $t$, computed numerically by choosing a sufficient number of steps $N_t$ until convergence is reached. 

This formulation significantly reduces the computational cost of implementing the kernel in Eq.~\eqref{eq:wagner_kernel}, for a separable target bispectrum. While a direct implementation, due to the non-separable denominator, would require an $\mathcal{O}(N_g^6)$ operation, the above expression allows for a much more efficient evaluation, scaling as $\mathcal{O}(N_tN_i\,N_g^3\,{\rm log}N_g)$ .

\section{Implementation and Code Structure}

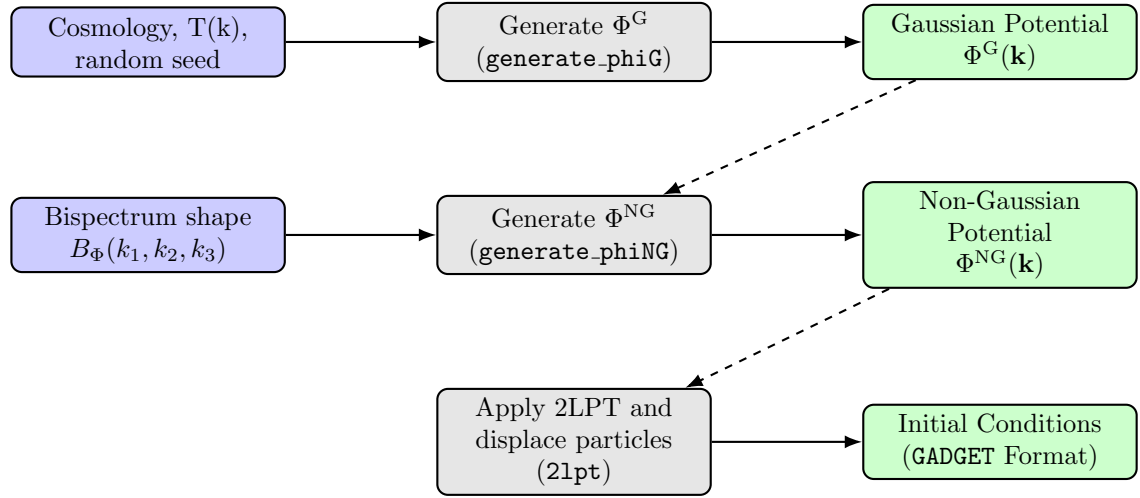
\begin{figure}[h]
    \centering
    \begin{tikzpicture}[
        node distance=6cm and 2cm,
        every node/.style={draw, text width=3.4cm, align=center, rounded corners},
        every path/.style={draw, -{Latex}, thick},
        dashedpath/.style={draw, dashed, -{Latex}},
        textnode/.style={align=center, font=\bfseries}  
    ]

    \node (phiG) [draw, rectangle, fill=gray!20] {Generate $\Phi^{\rm G}$ \\ (\texttt{generate\_phiG})};
    \node (phiNG) [draw, rectangle, fill=gray!20, below=1.5cm of phiG] {Generate $\Phi^{\rm NG}$ \\ (\texttt{generate\_phiNG})};
    \node (displace) [draw, rectangle, fill=gray!20, below=1.5cm of phiNG] {Apply 2LPT and displace particles \\ (\texttt{2lpt})};

    \node (param) [draw, rectangle, fill=blue!20, left=of phiG] {Cosmology, T(k), random seed};
    \node (bisp) [draw, rectangle, fill=blue!20, left=of phiNG] {Bispectrum shape $B_{\Phi}(k_1,k_2,k_3)$};

    \node (phiGout) [draw, rectangle, fill=green!20, right=of phiG] {Gaussian Potential \\ $\Phi^{\rm G}(\vk)$};
    \node (phiNGout) [draw, rectangle, fill=green!20, right=of phiNG] {Non-Gaussian Potential \\ $\Phi^{\rm NG}(\vk)$};
    \node (ICout) [draw, rectangle, fill=green!20, right=of displace] {Initial Conditions \\ (\texttt{GADGET} Format)};

    \draw (param) -- (phiG);
    \draw (bisp) -- (phiNG);
    \draw (phiG) -- (phiGout);
    \draw [dashedpath] (phiGout) -- (phiNG);
    \draw (phiNG) -- (phiNGout);
    \draw [dashedpath] (phiNGout) -- (displace);
    \draw (displace) -- (ICout);

    \end{tikzpicture}
    \caption{Workflow of the code. Each grey box represents a separate executable: \texttt{generate\_phiG}, \texttt{generate\_phiNG}, and \texttt{2lpt}. Arrows indicate the flow of information, with external inputs in blue and outputs in green. Dashed arrows indicate outputs that are reused by the executables.}
    \label{fig:workflow}
\end{figure}

The code presented here is an actualisation and optimization of the code used in Ref.~\cite{Wagner:2011wx}. It is written in C in a modular way and is MPI parallelized. It consists of three main executables, which we briefly describe below. Figure~\ref{fig:workflow} summarizes the structure of the code.

\subsection*{Gaussian potential contribution: \texttt{generate\_phiG}}
Given a set of cosmological parameters and the matter transfer function $T(k)$ (obtained e.g., from \texttt{CAMB} \cite{Lewis:1999bs}), the linear matter power spectrum $P(k)$ is computed. The power spectrum normalization can be specified either through $A_s$ or via $\sigma_8$, both supported as input options in the code.\\
The linear matter density field $\delta(\vk)=\sqrt{\frac{P(k)}{2}}(X + iY)$ is then generated in a cubic box of side $L$ with $N_g^3$ grid points. At each grid point, $X$ and $Y$ are extracted from a unit Gaussian distribution, using an input random seed for initializing the random number generator.\\
Finally, the Gaussian primordial potential $\Phi^{\rm G}(\vk)$ is obtained from $\delta(\vk)$ through the Poisson equation
\begin{equation}\label{eq:poisson}
\Phi(\vk)\equiv{\cal M}^{-1}(k,z)\delta(\vk)=\frac{3\Omega_mH_0^2}{2c^2k^2T(k)D(z)}\delta(\vk).
\end{equation}
\subsection*{Non-Gaussian potential contribution: \texttt{generate\_phiNG}}
The non-Gaussian contribution $\Phi^{\rm NG}(\vk)$ is generated  from the input $\Phi^{\rm G}(\vk)$ and the user-defined bispectrum, following Eq.~\eqref{eq:phi_conv_schwinger}. The input bispectrum $B_{\Phi}(k_1,k_2,k_3)$ must be factorized as in Eq.~\eqref{eq:sep_bk}.\\
The computation of $\Phi^{\rm NG}(\mathbf{k})$ involves a number of convolutions, performed using FFTs. The total number of FFTs required is:

\begin{equation}
    N_{\rm FFT} = N_i \times N_t,
\end{equation}

where $N_i$ is the number of terms which define the input bispectrum in Eq.~\eqref{eq:sep_bk}, while $N_t$ is the number of integration steps chosen for the discretization of the $dt$ integral in Eq.~\eqref{eq:phi_conv_schwinger}. The computational cost is therefore $\mathcal{O}(N_t N_i N_g^3 \log N_g)$.

Due to the structure in Eq.~\eqref{eq:phi_conv_schwinger}, the computation of $\Phi^{\rm NG}(\vk)$  generates modes corresponding to $\mathbf{k} + \mathbf{k}'$, which may lie outside the Nyquist range of the grid. In our current implementation, we do not apply any explicit treatment to these out-of-range modes. As discussed in \cite{Michaux2021, Adame2025}, a more refined approach to suppress aliasing is the so-called Orszag 3/2 rule \cite{Orszag1971}, which involves zero-padding the Fourier grid before performing nonlinear operations. However, Ref.~\cite{Adame2025} showed that for the case of local PNG, the aliasing signal is significantly suppressed at $z=0$, making its impact negligible for the level of accuracy required by current simulations. Given the substantial memory overhead introduced by the Orszag 3/2 rule, we do not implement it by default, though we will include it as an optional feature in future releases.

\subsection*{Initial conditions for particle positions and velocities: \texttt{2lpt}}
The primordial potential $\Phi(\vk)=\Phi^{\rm G}(\vk)+\fnl \Phi^{\rm NG}(\vk)$ is computed from the Gaussian and non-Gaussian contributions and the input $\fnl$. The Poisson equation \eqref{eq:poisson} is then used to obtain $\delta(\vk,z)$ at the simulation starting redshift $z$. $N_p^3$ particles are initially placed at the grid points $\mathbf{q}$, forming a regular lattice. To properly account for gravitational evolution from the early times up to the starting redshift $z$, their positions and velocities are computed using Lagrangian perturbation theory (LPT). The gravitational dynamics is captured by the Lagrangian displacement field $\mathbf{\Psi}(\mathbf{q})$, such that the particle positions and velocities are given by: 
\begin{align}
\mathbf{x}(\mathbf{q}, z) &= \mathbf{q} + \mathbf{\Psi}(\mathbf{q}, z)\\
\mathbf{v}(\mathbf{q}, z) &= \dot{\mathbf{\Psi}}(\mathbf{q}, z)
\end{align}

and $\mathbf{\Psi}$ is expanded at n-th order, with each expansion term being computed from derivatives of the density $\delta$ \cite{Peebles_1980,Bouchet1992,bernardeau2002}. By default, our implementation includes second-order LPT (2LPT). Currently, particles are initialized on a regular grid, and \texttt{GENGARS} does not support glass-like configurations. Extending the code to implement glass initial conditions will be considered in future updates. Finally, the initial conditions are written in \texttt{GADGET} format \cite{Springel:2005}.\\ 

The modular structure of the code allows flexibility in how initial conditions are generated. Once $\Phi^{\rm NG}$ is computed, the initial conditions can be straightforwardly obtained for different values of $f_{\rm NL}$ without re-computing $\Phi^{\rm NG}$. Furthermore, this implementation ensures compatibility with different prescriptions for particle displacement \cite{Michaux2021}, as well as different simulation codes beyond \texttt{GADGET} \cite{Potter2017,Springel2021,Feng2016,Howlett2015}.

\section{Comparison with \texttt{2LPT-PNG}: initial conditions}

In this section, for the common PNG templates (local, equilateral and orthogonal) in Eqs.~\eqref{eq:b_loc}-\eqref{eq:b_ort}, we compare the summary statistics of the primordial gravitational potential generated with \texttt{GENGARS} with those obtained using \texttt{2LPT-PNG}, which is based on the formulation by Ref.~\cite{Scoccimarro12}. The fundamental difference between the two approaches lies in the definition of the kernels $W(k_1,k_2,k_3)$ used to generate the primordial gravitational potential in Eq.~\eqref{eq:generic_phi}.

As mentioned in Sec.~\ref{subsec:kernel_choice},  Ref.~\cite{Scoccimarro12} solves a different inverse problem in Eq.~\eqref{eq:kernel_eq} for each different template.

The kernels in \texttt{2LPT-PNG} are explicitly constructed in a separable form, allowing the use of FFTs to efficiently generate non-Gaussian fields, as in Eq.~\eqref{eq:phi_conv}.

Finally, they are tuned in order to regularize the scaling of the non-Gaussian contribution to the primordial power spectrum in Eq.~\eqref{eq:ng_pk_contribution}. For local PNG both codes use $W^{\rm loc}\equiv 1$, while the kernels differ for the equilateral and orthogonal templates. Understanding the impact of the different choices, comparing the two implementations, is crucial to ensure accurate and controlled initial conditions for cosmological simulations with PNG.

We first generate a common Gaussian primordial potential $\Phi^{\rm G}$ and then pass it to both codes to construct the non-Gaussian contributions $\Phi^{\rm NG}$ for local, equilateral, and orthogonal bispectrum templates. This ensures that any differences observed in the summary statistics arise solely from differences in kernel construction, rather than from variations in the initial Gaussian field. The initial conditions are generated in a box of side $L=1Gpc/h$ using a grid with $512^3$ points. We repeat this procedure for 10 independent realizations of the Gaussian field.

We show our comparison with \texttt{2LPT-PNG} in terms of the primordial power spectrum and bispectrum. These statistics are computed using \texttt{Pylians3}\footnote{\url{https://github.com/franciscovillaescusa/Pylians3}}. We measure the primordial potential statistics on a $512^3$ grid constructed using the Cloud-in-Cell (CIC) assignment scheme, and correct the resulting power spectra and bispectra by deconvolving the CIC window function.

\subsection{Primordial power spectrum}

As discussed in Sec.~\ref{subsec:kernel_choice}, the primordial power spectrum in initial conditions with PNG includes a non-Gaussian contribution, given by $P_\Phi^{\rm NG}(k)$ in Eq.~\eqref{eq:ng_pk_contribution}, and an additional mixed term $\langle \Phi^{\rm G}(\mathbf{k}) \Phi^{\rm NG}(\mathbf{k'}) \rangle$, which is expected to vanish theoretically but can be non-zero in practice due to the finite number of modes in the simulation volume.

\begin{figure}[h]
    \centering
    \includegraphics[width=\textwidth]{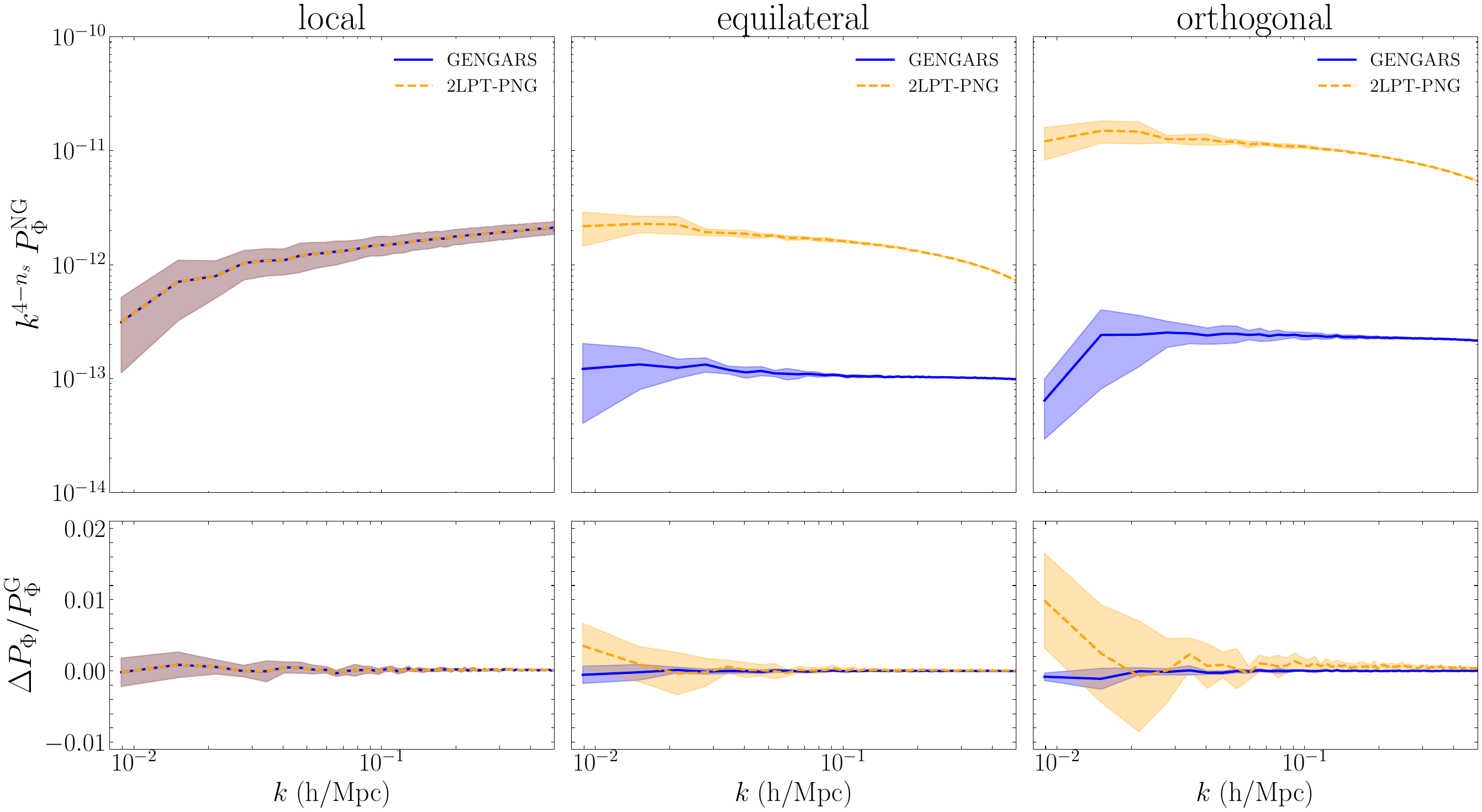}
    \caption{Primordial power spectrum corrections due to PNG, computed from ten realizations with $f_{\rm NL} = 100$, for the local, equilateral, and orthogonal templates. In the top panels, we show the non-Gaussian contribution as $k^{n_s - 4} P_\Phi^{\rm NG}(k)$, which isolates its scaling compared to the Gaussian part. In the bottom panels, we report the total fluctuation relative to the Gaussian primordial power spectrum, $(P_{\Phi} - P_{\Phi}^{\rm G}) / P_{\Phi}^{\rm G}$. The shaded areas represent the error on the mean, computed over 10 independent realizations.}
    \label{fig:pk_comp}
\end{figure}

In Fig.~\ref{fig:pk_comp}, we show the non-Gaussian corrections to the primordial power spectrum for different bispectrum templates. In the top panel, we plot $k^{n_s - 4} P_\Phi^{\rm NG}(k)$, which helps visualize whether the scaling of this term would alter the large-scale behavior of the total power spectrum. The bottom panel shows the relative difference between the total measured power spectrum and the Gaussian one, $(P_{\Phi} - P_{\Phi}^{\rm G}) / P_{\Phi}^{\rm G}$. Both panels report the mean and the error on the mean computed over 10 independent realizations, with $f_{\rm NL} = 100$.

For the local shape, the two implementations yield identical results, as expected since both use the same kernel $W=1$. In all other cases, we find that \texttt{GENGARS} produces a systematically smaller non-Gaussian contribution to the power spectrum. This is a consequence of the kernel suppression in the squeezed limit introduced by the denominator in Eq.~\eqref{eq:wagner_kernel}, as discussed above. Both implementations produce corrections scaling at most as $k^{n_s - 4}$, thus preserving the large-scale tilt of the primordial spectrum. In particular, for the orthogonal case, the correction in \texttt{GENGARS} diverges slower than $k^{n_s - 4}$, leading to a subdominant effect on large scales.

In terms of amplitude, the contribution $P_\Phi^{\rm NG}$ from \texttt{GENGARS} is consistently smaller than that from \texttt{2LPT-PNG}, and also smaller than the local PNG case for the other templates. As mentioned above, this reflects the fact that the denominator of our kernel also regularizes the amplitude, determined by the limit $k' \to 0$ in Eq.~\eqref{eq:ng_pk_contribution}. In contrast, the kernels used in \texttt{2LPT-PNG} do not fully control the $k'$ dependence, which leads to an enhanced amplitude for the equilateral and orthogonal shapes. Considering the amplitude of the Gaussian power spectrum $P_\Phi^{\rm G}(k) \sim 10^{-9}$, the non-Gaussian correction from \texttt{GENGARS} remains at least three orders of magnitude smaller for $f_{\rm NL} = 100$. For \texttt{2LPT-PNG}, however, the orthogonal case introduces a correction at the 1\% level, which implies that for $f_{\rm NL} = 1000$, $P_\Phi^{\rm NG}(k)$ could become comparable to the Gaussian part, potentially altering the overall slope.

In the bottom panels, the variance across realizations reflects contributions from both the non-Gaussian power spectrum term $P_\Phi^{\rm NG}$ and the mixed term $\langle \Phi^{\rm G} \Phi^{\rm NG} \rangle$ in Eq.~\eqref{eq:pk_phi_terms}. Again, due to the structure of our kernel, the variance is suppressed with respect to \texttt{2LPT-PNG} and is stable across the different templates.

\subsection{Primordial bispectrum}

\begin{figure}[h]
    \centering
    \includegraphics[width=\textwidth]{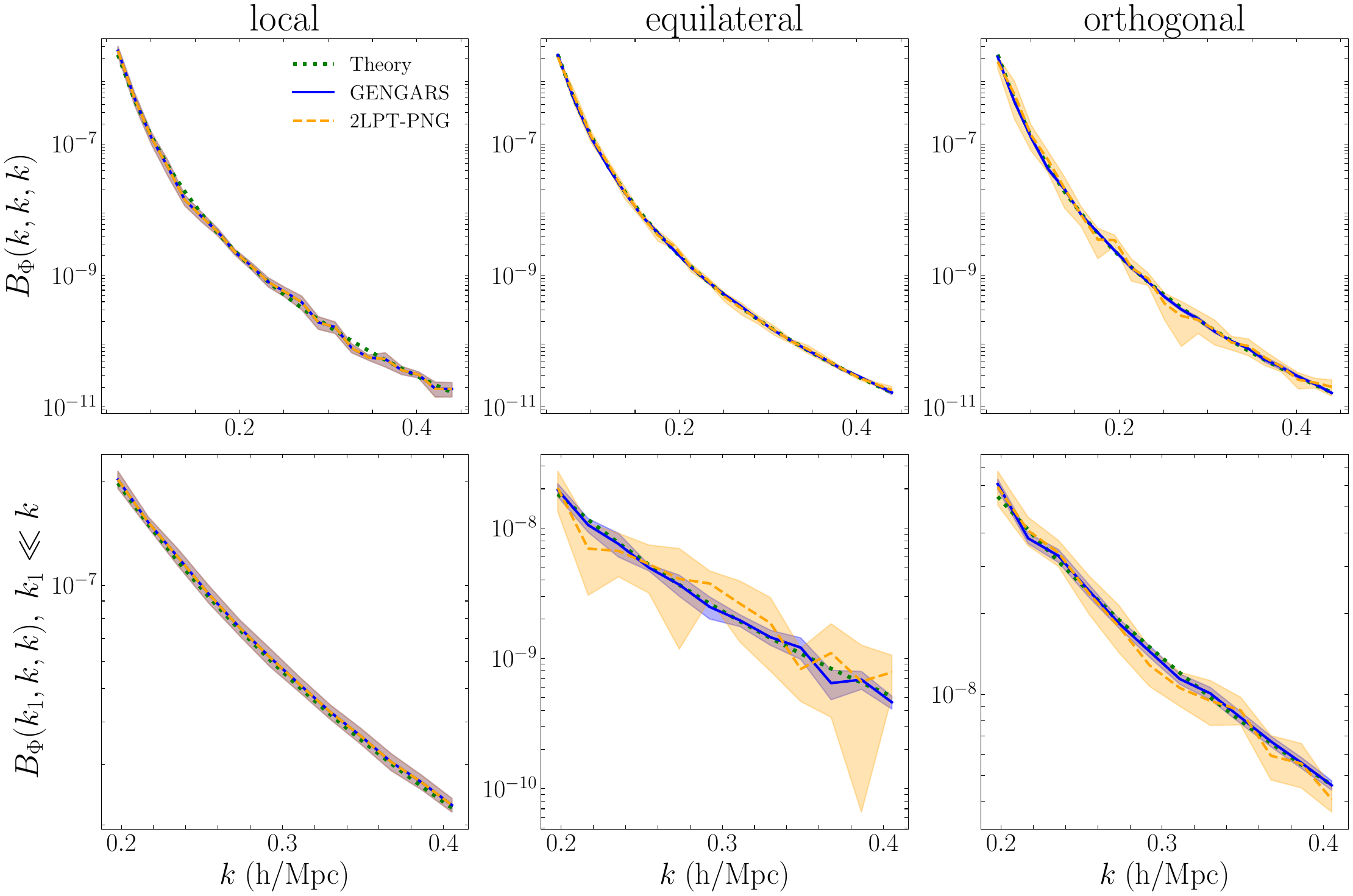}
    \caption{Primordial bispectrum of the gravitational potential $B_\Phi$ measured in the equilateral (top) and squeezed (bottom) configurations for the local, equilateral, and orthogonal templates, with $f_{\rm NL}=100$. For the squeezed bispectrum, we fix $k_1=3k_F \ll k_2=k_3\equiv k$. Solid blue lines refer to \texttt{GENGARS}, dashed orange lines to \texttt{2LPT-PNG}, and green dotted lines show the target input bispectrum. Each curve shows the mean over 10 realizations, while the shaded regions represent the error on the mean, i.e. divided by $\sqrt{10}$.}
    \label{fig:bk_comp}
\end{figure}

In Fig.~\ref{fig:bk_comp}, we show the primordial bispectrum $B_\Phi$ evaluated in the equilateral (top row) and squeezed (bottom row) configurations, for the local, equilateral, and orthogonal templates. For the squeezed bispectrum, we fix $k_1=3k_F \ll k_2=k_3$, with $k_F=2\pi /L$ being the fundamental mode. Each panel compares the results from the \texttt{GENGARS} and \texttt{2LPT-PNG} implementations, together with the target bispectrum.

Both methods successfully reproduce the expected mean bispectrum shape, confirming that the desired three-point function is correctly generated in all cases. However, there are differences in the variance across realizations, especially in the squeezed configurations. In particular, the scatter is consistently smaller in \texttt{GENGARS}, which may be attributed to the suppression induced by the denominator of the kernel in Eq.~\eqref{eq:wagner_kernel}. This structure effectively suppresses contributions from higher-order terms, thereby reducing the noise. Possible consequences of these differences on the statistics of the evolved field  are discussed in Sec.~\ref{sec:evolved}.

Despite this reduced variance, it is important to stress that our algorithm only controls the three-point function. The bispectrum variance is determined by the six-point function of the field, over which we have no direct control. Thus, while a smaller scatter may appear favorable, it is not necessarily a sign of improved accuracy beyond the bispectrum itself.

The increased variance observed in the squeezed bispectrum of the equilateral shape is expected: the signal in this configuration is intrinsically suppressed, as also discussed in \cite{Coulton:2022rir}. Consequently, cosmic variance plays a dominant role, making fluctuations more prominent across realizations.

\subsection{Computational time}
In Table~\ref{tab:comp_time}, we report the computational time required to generate $\Phi^{\rm NG}$ from a given Gaussian field $\Phi^{\rm G}$, for different grid sizes $N_{\rm g}$ and numbers of terms $N_i$ in the bispectrum decomposition. The timings correspond to runs performed on a total of 64 CPUs across 2 nodes.

The approach of Ref.~\cite{Wagner:2011wx} is based on the brute-force evaluation of Eq.~\eqref{eq:generic_phi}, which scales as $\mathcal{O}(N_{\rm g}^6)$ and depends only on grid resolution. As expected, this makes the method computationally prohibitive for large grids, with run times extending to several days for $N_{\rm g}=512$.

Our separable implementation of the reduced bispectrum kernel exploits FFTs and reduces the scaling to $\mathcal{O}(N_t N_i N_{\rm g}^3 \log N_{\rm g})$. The overhead with respect to \texttt{2LPT-PNG} is due to the numerical integration over the Schwinger parameter $t$. The number of steps $N_t$ required for convergence has been tested by comparing our results against the brute-force computation of Eq.~\eqref{eq:generic_phi} for the kernel in Eq.~\eqref{eq:wagner_kernel}. We typically find that $N_t \sim 300$--400 steps are sufficient, depending on the number of grid points per side, $N_g$. As a result, while \texttt{GENGARS} is computationally more expensive than \texttt{2LPT-PNG}, it remains several orders of magnitude faster than the original implementation in Ref.~\cite{Wagner:2011wx}.

\begin{table}[H]
    \centering
    \begin{tabular}{|c|c|c|c|c|}
    \hline
    $N_i$ & $N_{\rm g}$ & Ref.~\cite{Wagner:2011wx} & \texttt{2LPT-PNG} & \texttt{GENGARS}\\
    \hline
    4 & 256 & 5 hrs & 2 s & 4 min \\
    4 & 512 & 13 days & 13 s & 30 min \\
    10 & 256 & 5 hrs & 4 s & 9 min \\
    10 & 512 & 13 days & 27 s & 75 min \\
    \hline
    \end{tabular}
 \caption{Time required to generate $\Phi^{\rm NG}$ on a total of 64 CPUs across 2 nodes, for different grid sizes $N_{\rm g}$ and number of terms $N_i$ in the bispectrum decomposition.}
    \label{tab:comp_time}
\end{table} 

\section{Application to non-standard bispectrum shapes: oscillatory feature example}

While the previous sections have focused on the standard templates (local, equilateral, and orthogonal), a primary advantage of our implementation is its capability to generate initial conditions for arbitrary separable bispectrum shapes without the need to find and implement the generating kernel, solution to Eq.~\eqref{eq:kernel_eq}. To illustrate this flexibility, we now briefly consider a model featuring an oscillatory bispectrum, which can arise in inflationary scenarios with sharp features in the inflaton potential \cite{chen2010, Euclid2020pxf}.

We consider the following bispectrum template, characterized by oscillations with frequency $\omega$ and phase $\phi$:

\begin{equation}\label{eq:oscillatory_bispectrum}
B_{\Phi}^{\mathrm{osc}}(k_1,k_2,k_3) = 6 f_{\mathrm{NL}}^{\mathrm{osc}}\left[P_{\Phi}(k_1)P_{\Phi}(k_2)P_{\Phi}(k_3) \right]^{2/3}\sin\left[\omega(k_1 + k_2 + k_3) + \phi\right].
\end{equation}

\begin{figure}[H]
\centering
\includegraphics[width=0.7\textwidth]{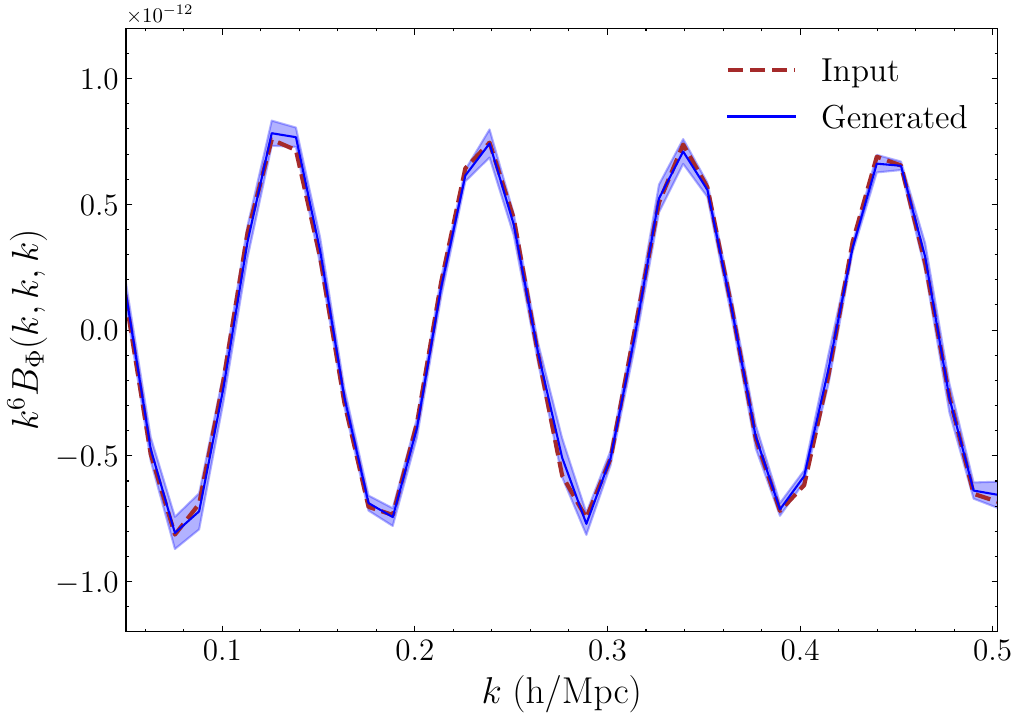}
\caption{Comparison of the primordial bispectrum generated with \texttt{GENGARS} (solid blue line, averaged over 5 realizations, with the shaded area representing scatter) against the input (brown dashed line) from the oscillatory bispectrum template in Eq.~\eqref{eq:oscillatory_bispectrum}, with $\omega=20$ and $\phi=0$. Results are shown for equilateral configurations ($k_1=k_2=k_3=k$).}
\label{fig:oscillatory_example}
\end{figure}

Figure~\ref{fig:oscillatory_example} shows the primordial bispectrum generated with \texttt{GENGARS} for this oscillatory template with $\omega=20$ and $\phi=0$, computed on equilateral configurations. The plot compares the generated bispectrum (averaged over 5 independent realizations) with the input theoretical bispectrum from Eq.~\eqref{eq:oscillatory_bispectrum}, highlighting the excellent accuracy achieved by our implementation. The shaded area indicates the scatter across realizations, which remains consistently small.

This example shows how \texttt{GENGARS} allows us to reproduce arbitrary shapes without requiring any approximation or analytical treatment beyond specifying the functional form. Such capability broadens the range of inflationary models that can be tested through cosmological N-body simulations, providing a versatile tool for investigating the impact and detectability of primordial non-Gaussianity.

\section{Comparison with \texttt{2LPT-PNG}: evolved dark matter and halo field}
\label{sec:evolved}
In this Section, we compare the late-time matter and halo statistics obtained from N-body simulations initialized with non-Gaussian initial conditions generated using \texttt{GENGARS} and\texttt{2LPT-PNG}. To do so, we analyze the first 10 realizations of the QuijotePNG \cite{Coulton:2022qbc,Coulton:2022rir,Jung2023} suite for equilateral and orthogonal templates with $f_{\rm NL} = \pm100$. These simulations use \texttt{2LPT-PNG} for generating non-Gaussian initial conditions and serve as our baseline for comparison. We focus on equilateral and orthogonal PNG since the local template is implemented identically in both codes, leading to identical initial conditions and thus no expected differences in the evolved field. 

To ensure a consistent and controlled comparison, we generate the Gaussian potentials $\Phi^{\rm G}$ by running \texttt{2LPT-PNG} with the same random seeds used in the fiducial Quijote simulations\cite{villaescusa-navarro_quijote_2020} ($f_{\rm NL}=0$). The Gaussian potentials are then passed to \texttt{GENGARS} to compute the corresponding non-Gaussian contributions $\Phi^{\rm NG}$. Particles are then displaced using 2LPT and evolved to redshift $z=0$ with \texttt{GADGET-3}, using the same cosmological parameters and simulation settings as in QuijotePNG. By construction, this setup isolates differences arising exclusively from the PNG kernel implementation.

All simulations assume a box of size $L = 1\, h^{-1}\mathrm{Gpc}$ with $512^3$ dark matter particles, and initial conditions are generated at $z = 127$ using a grid with $1024^3$ points. Halos are identified at $z=0$ using a Friends-of-Friends (FoF) algorithm with a linking length of $b = 0.2$.

In the following figures, we present the dark matter power spectra and bispectra, as well as the halo mass function for the different cases, averaged over the 10 realizations considered.

\subsection{Matter power spectrum}
\begin{figure}[H]
    \centering
    \includegraphics[width=\textwidth]{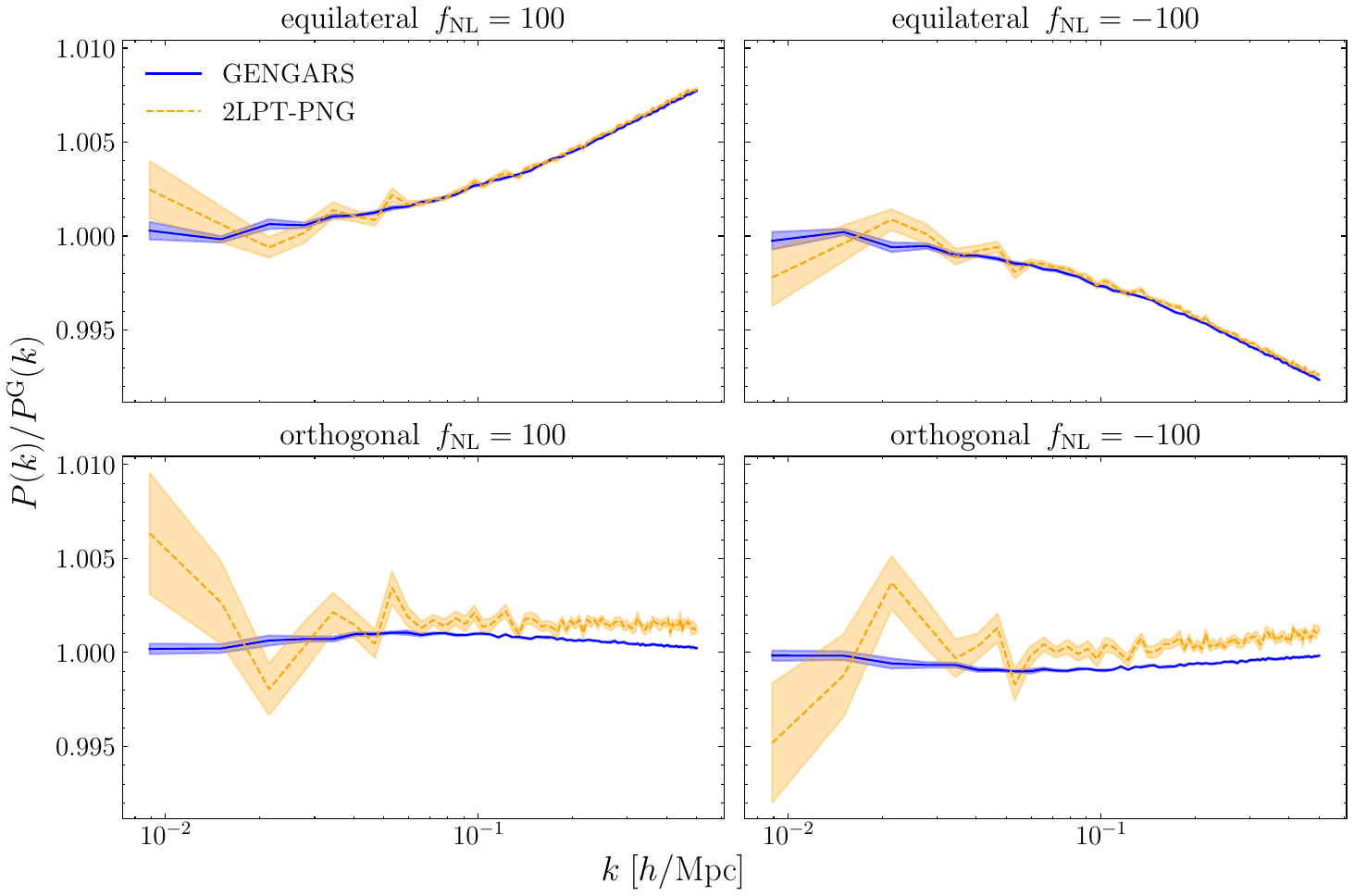}

\caption{Ratio between the matter power spectrum measured at $z = 0$ in simulations with PNG and in the corresponding Gaussian simulations, for equilateral and orthogonal templates with $f_{\rm NL} = \pm 100$. Solid blue lines correspond to \texttt{GENGARS}, dashed orange lines to \texttt{2LPT-PNG}. The shaded bands represent the error on the mean over 10 realizations.}

    \label{fig:pk_grid}
\end{figure}

We begin by comparing the matter power spectra at $z=0$ for the four PNG scenarios considered in this work.
In Fig.~\ref{fig:pk_grid} we present the results for the ratio between the power spectrum measured from the non-Gaussian simulations and their Gaussian counterpart. For each scenario, we compute the mean and the error on the mean over 10 realizations.

We focus on the ratio $P(k)/P^{\rm G}(k)$ since deviations from the Gaussian case remain below the percent level. The results at $z=0$ follow from the behavior observed at the level of the initial conditions. In particular, the equilateral and orthogonal cases with $f_{\rm NL}=100$ (left panels) are consistent with the evolution of the spectra in the central and right bottom panels of Fig.~\ref{fig:pk_comp}. The large-scale contribution related to $P_{\Phi}^{\rm NG}$, dependent on the kernel choice, propagates linearly to the large-scale evolved power spectrum.

Coherently with the initial conditions results, \texttt{2LPT-PNG} introduces a larger contribution, especially for the orthogonal case. Although it remains below the percent level, we stress that this contribution is not physical—it originates from the structure of the kernel implemented in \texttt{2LPT-PNG}. Such large-scale contributions are below the 0.1\% level in the \texttt{GENGARS} results, confirming that the suppression in the squeezed limit introduced by our kernel reduces these spurious effects. In the same way, noise fluctuations are also suppressed, similarly to what we observed in the initial conditions.

On smaller scales, for the orthogonal case, we observe discrepancies between the two implementations, due to the nonlinear evolution of the differences in the primordial kernels. These differences could become relevant in analyses that include small-scale information. However, assessing their impact, or determining which prescription is more accurate in this regime, goes beyond the scope of this work and is left for future study.

\subsection{Matter bispectrum}

Since the contribution from gravitational non-Gaussianity dominates at $z=0$, we isolate the signal due to PNG by subtracting the matter bispectrum of the Gaussian counterpart from that of the PNG simulations. We show the equilateral bispectrum $\Delta B(k,k,k)$ in the range $[k_F,0.5$h/Mpc], with $k_F=2\pi /L$ fundamental mode. For the squeezed configurations, we fix $k_1\equiv3k_F$ and vary $k_2=k_3\equiv k$. We consider $k$ ranging from $10k_F$ up to $70 k_F$ to ensure sufficiently squeezed triangle shapes.

\begin{figure}[H]
    \centering

    \begin{minipage}[t]{0.48\textwidth}
        \centering
        \includegraphics[width=\linewidth]{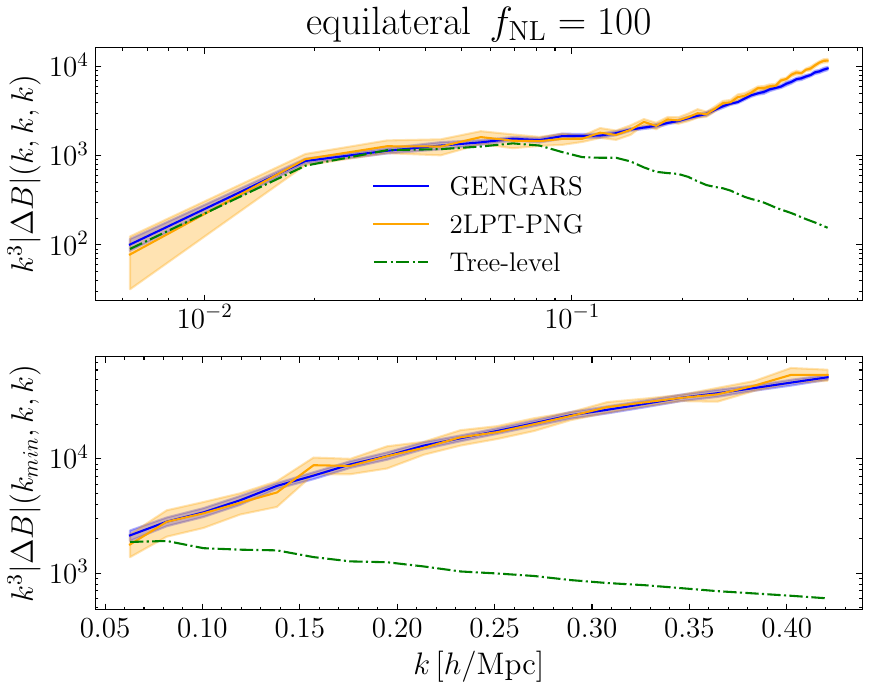}
    \end{minipage}
    \hfill
    \begin{minipage}[t]{0.48\textwidth}
        \centering
        \includegraphics[width=\linewidth]{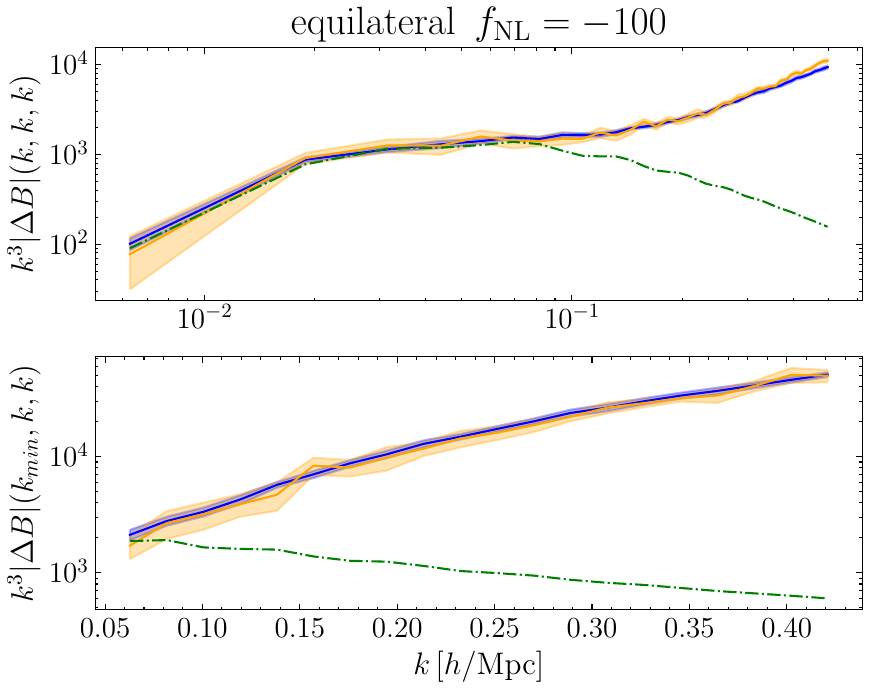}
    \end{minipage}
    \vspace{0.5cm}

    \begin{minipage}[t]{0.48\textwidth}
        \centering
        \includegraphics[width=\linewidth]{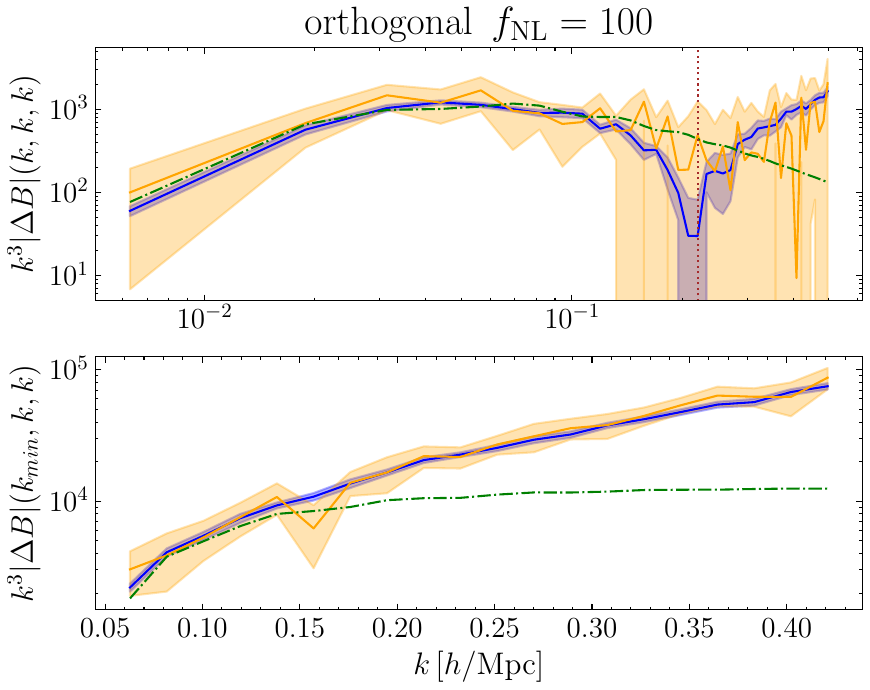}
    \end{minipage}
    \hfill
    \begin{minipage}[t]{0.48\textwidth}
        \centering
        \includegraphics[width=\linewidth]{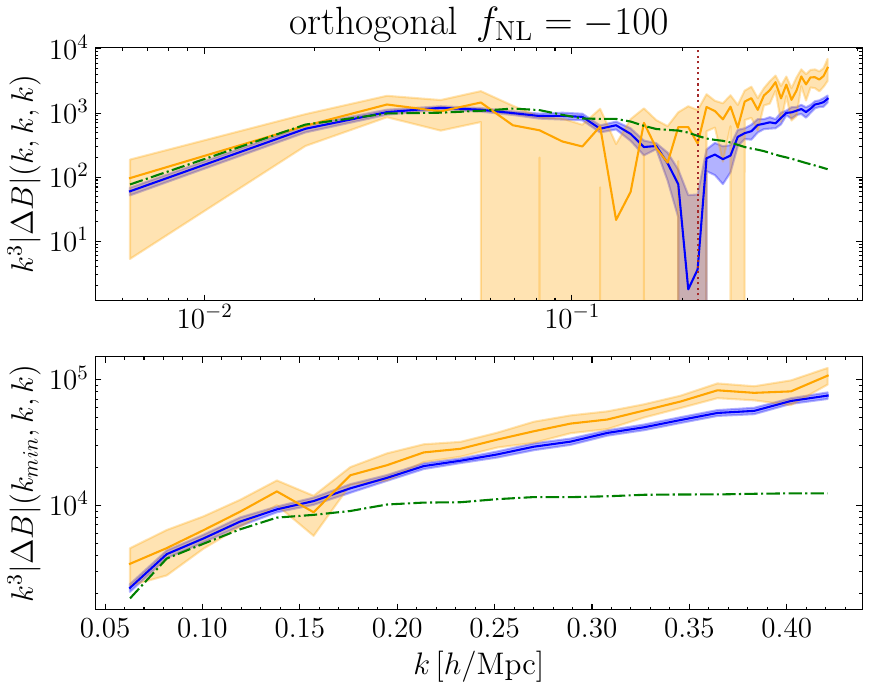}
    \end{minipage}

\caption{Comparison of the matter bispectrum at $z=0$ in equilateral (top panels) and squeezed (bottom panels) configurations, for simulations initialized with \texttt{GENGARS} (blue) and \texttt{2LPT-PNG} (orange). Each curve shows the mean over 10 realizations of the difference $B^{\rm NG} - B^{\rm G}$, where $B^{\rm NG}$ is the bispectrum measured in simulations with PNG, and $B^{\rm G}$ is the corresponding bispectrum from Gaussian simulations. Shaded regions represent the error on the mean, computed as the standard deviation across realizations divided by $\sqrt{10}$. The tree-level prediction is shown for comparison (green dashed line), and agrees with simulations on large linear scales ($k \lesssim 0.1\,h/{\rm Mpc}$). The results are plotted in absolute value, in the orthogonal case the sign flips at small scales, marked by the brown dotted vertical lines.}
    \label{fig:bk_z0_grid}
\end{figure}
\newpage
We compare the results obtained using \texttt{GENGARS} and \texttt{2LPT-PNG}, and include the tree-level theoretical prediction computed as
\begin{equation}
    B(k_1,k_2,k_3) = \mathcal M(k_1) \mathcal M(k_2) \mathcal M(k_3) B_{\Phi}(k_1,k_2,k_3),
\end{equation}
where $ \mathcal M(k)$ is the Poisson kernel defined in Eq.~\eqref{eq:poisson}. The results are shown in Fig.~\ref{fig:bk_z0_grid}. The tree-level prediction agrees well with the simulations on large scales ($k \lesssim 0.1\,h/{\rm Mpc}$), but breaks down at smaller scales due to non-linear evolution.

For visual purposes, we plot the bispectrum on a logarithmic scale and show the absolute value of $k^3\Delta B$, since the sign of the signal is not preserved across all configurations. In the equilateral case with negative $f_{\rm NL}$, the bispectrum is negative across all scales, while for the orthogonal template the sign changes at small scales. These sign reversals are indicated by brown dotted vertical lines in the equilateral panels. They can be attributed to higher-order contributions, arising from the coupling between PNG and gravitational evolution.

On large linear scales, both implementations give consistent results. Differences emerge at smaller scales due to the distinct non-linear evolution of the initial conditions, which are affected by the different kernel structures. However, it is not straightforward to assess which implementation is more accurate in this regime. Doing so would require extending the theoretical prediction beyond tree level including the 1-loop bispectrum, which lies beyond the scope of this work.

We also observe that \texttt{GENGARS} exhibits a generally smaller variance across realizations compared to \texttt{2LPT-PNG}. This trend is consistent with what we found for the bispectrum at the initial conditions and may be attributed to the suppression induced by the denominator in Eq.~\eqref{eq:wagner_kernel}. While the kernel is designed to reproduce the correct bispectrum, we have no control over higher-order statistics such as the six-point function. As such, the reduced variance is an interesting feature but does not necessarily imply improved accuracy in the non-linear regime.

\subsection{Halo mass function}
\begin{figure}[H]
    \centering
    \includegraphics[width=\textwidth]{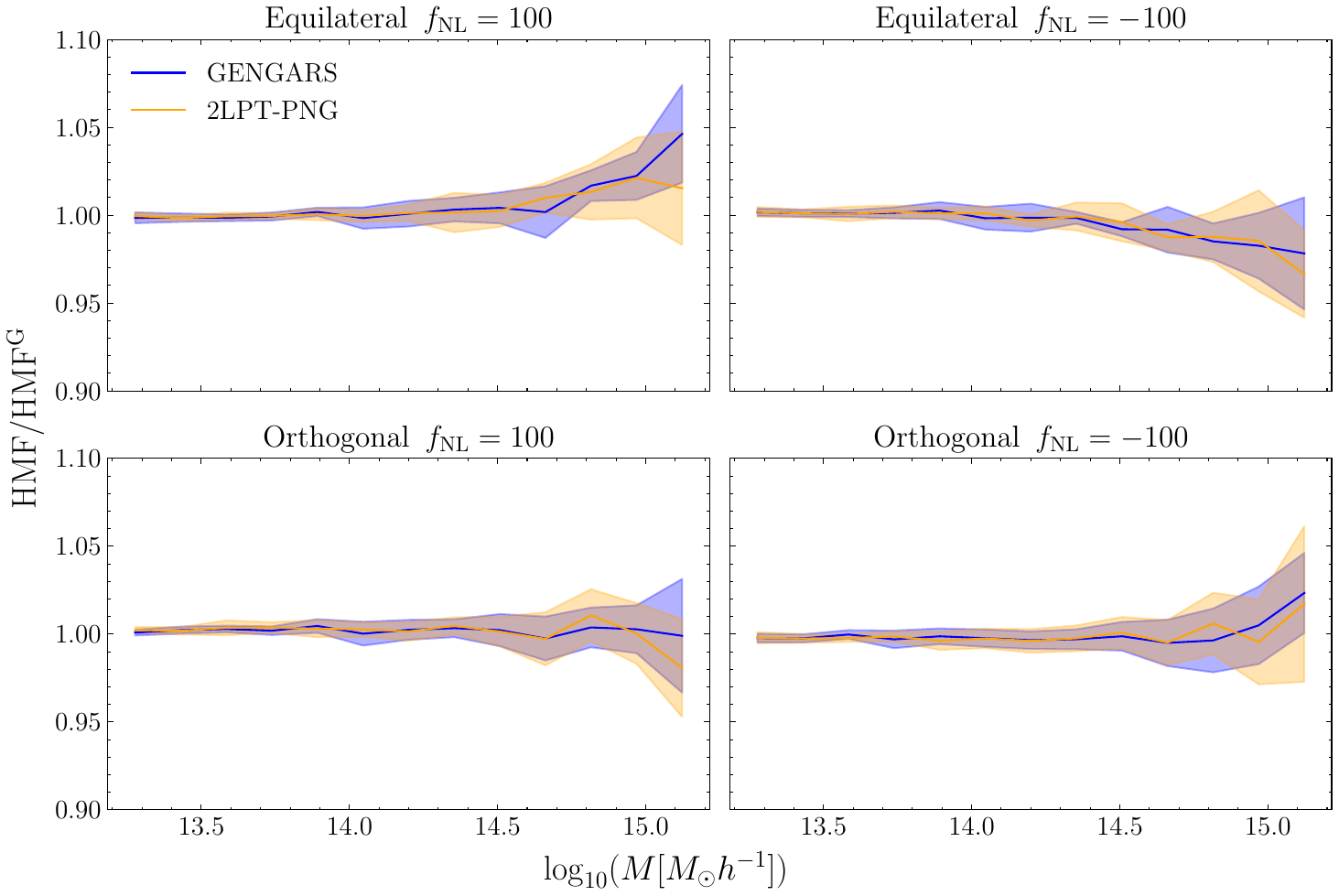}
    \caption{Ratio between the halo mass function (HMF) at $z = 0$ measured in simulations with PNG and the corresponding Gaussian HMF ($\mathrm{HMF}^\mathrm{G}$). Each panel corresponds to a different PNG configuration (equilateral and orthogonal with $f_{\rm NL} = \pm 100$). Blue lines refer to \texttt{GENGARS}, orange lines to \texttt{2LPT-PNG}. The shaded bands represent the $1\sigma$ dispersion across 10 realizations.}
\label{fig:hmf_grid}
\end{figure}

We now turn to the halo mass function (HMF) at redshift $z = 0$. In analogy with the power spectrum, in Fig.~\ref{fig:hmf_grid} we show the HMF for the four PNG scenarios considered, and plot the ratio between the non-Gaussian and Gaussian HMF. For each scenario, we bin halos in logarithmic mass bins and count the number of halos in each bin, using total of 14 bins in the range $\log_{10}(M)=[13.2, 15.2]$, with $M$ in units of $M_\odot h^{-1}$.

Differences between the PNG and Gaussian cases remain small, typically below $5\%$, and are more pronounced at high halo masses. This is expected since PNG primarily affects the abundance of rare, massive halos due to its impact on the tails of the initial density distribution \cite{Jung2023}.
At large masses, where the effect is most visible, statistical uncertainties increase due to the low number of halos. In all four cases the two implementations remain fully consistent within the $1\sigma$ dispersion. Unlike what we observed for the dark matter power spectrum and bispectrum, the variance of the HMF ratio appears comparable between the \texttt{GENGARS} and \texttt{2LPT-PNG}. This likely happens because halo formation and finding are highly nonlinear and stochastic processes, and the variance in halo counts is dominated by shot noise, rather than by differences in the PNG initial conditions prescriptions.

\section{Conclusions}
Cosmological simulations represent a crucial tool to study the signatures of PNG on large-scale structure and assess their detectability, providing a window into the nature of inflation.
In order to run N-body simulations with PNG, one needs to generate initial conditions in which the primordial gravitational potential $\Phi$ accurately reflects the desired bispectrum $B$.
This task involves solving an inverse problem: given a target bispectrum, the corresponding generating kernel $W$ must be determined. However, the solution of this problem is not unique -- different kernels can reproduce the same bispectrum. In practice, this ambiguity can be constrained by requiring that: (i) the kernel is separable, allowing a fast computation through FFTs; (ii) it should not spoil the primordial power spectrum on large scales.

In this work, we presented \texttt{GENGARS} (\textbf{GE}nerator of \textbf{N}on-\textbf{G}aussian \textbf{AR}bitrary \textbf{S}hapes), a framework that implements the reduced bispectrum kernel introduced by Wagner \& Verde \cite{Wagner:2011wx}.
This kernel fixes the solution to the inverse problem for arbitrary bispectrum shapes, without requiring any analytical treatment beyond specifying the functional form of the target bispectrum. Its implementation is made computationally efficient by exploiting the Schwinger parameterization, provided that the target bispectrum is specified in a separable form. Compared to the original brute-force implementation \cite{Wagner:2011wx}, this approach improves computational performance by orders of magnitude, making it feasible to generate initial conditions for arbitrary PNG models at high resolution.

We benchmarked our implementation against the well-established \texttt{2LPT-PNG} code \cite{Scoccimarro12,Coulton:2022qbc} across widely used bispectrum templates (local, equilateral, and orthogonal). While both methods reproduce the correct mean bispectrum, \texttt{GENGARS} tends to produce a smaller variance across realizations and a smaller spurious contribution to the primordial power spectrum. This is due to the suppression introduced by the denominator in the kernel, which regularizes its behavior in the squeezed limit. 

We further tested the impact of the different implementations on the evolved field by comparing matter and halo statistics at $z = 0$ from 10 realizations of N-body simulations initialized with \texttt{GENGARS} and \texttt{2LPT-PNG}. For the matter power spectrum, the spurious contribution present in the \texttt{2LPT-PNG} initial conditions translate into a large-scale excess of power (below 1\%), particularly for orthogonal PNG. In contrast, this unphysical contribution is absent in \texttt{GENGARS}, and noise fluctuations are also suppressed. These results highlight the accuracy and robustness of our prescription for generating non-Gaussian initial conditions. For the large-scale matter bispectra, instead, we find general agreement between the two implementations. At smaller scales, we observe small discrepancies between the two methods. These differences are likely due to the different higher-order statistics of the initial fields, which are not controlled by construction, as both methods are designed to match the bispectrum but not higher-order moments. Finally, for the halo mass function, both methods yield consistent results within sample variance, and no systematic differences are observed across the full mass range.

Although \texttt{GENGARS} allows for the accurate generation of arbitrary separable bispectra, the use of Schwinger parameterization introduces an overhead relative to \texttt{2LPT-PNG}. For standard templates, \texttt{2LPT-PNG} remains more efficient and is generally preferred for large production runs. However, the improved accuracy and lower variance of our method may still make \texttt{GENGARS} the better choice in applications where precision is critical. Further investigation is needed to assess whether the differences in the initial conditions could lead to biases in higher-order statistics on smaller scales.

Overall, \texttt{GENGARS} is particularly suited for generating non-Gaussian initial conditions for arbitrary or non-standard separable bispectrum shapes, enabling simulation-based studies of a broader class of inflationary models. While the current implementation is CPU-based, further improvements—such as GPU acceleration using frameworks like JAX—are left for future work. These developments could significantly reduce runtime and improve scalability, especially for high-resolution simulations.

\acknowledgments
EF acknowledges the support from ``la Caixa” Foundation (ID 100010434, code LCF/BQ/DI21/11860061).
Funding for this work was partially provided by project PGC2018-098866-B-I00 \\ MCIN/AEI/10.13039/501100011033 y FEDER “Una manera de hacer Europa”, the “Center of Excellence Maria de Maeztu 2020-2023” award to the ICCUB (CEX2019-000918-M funded by \\ MCIN/AEI/10.13039/501100011033) and  “Center of Excellence Maria de Maeztu" award to the ICCUB CEX2024-001451-M funded by MICIU/AEI/10.13039/501100011033.
LV acknowledges support of  European Union's Horizon 2020 research and innovation programme ERC (BePreSysE, grant agreement 725327). EF and BDW benefited from the hospitality of the program CoBALt, held at the Institut Pascal at Université Paris-Saclay with the support of the program “Investissements d’avenir” ANR-11-IDEX-0003-01.

\providecommand{\href}[2]{#2}\begingroup\raggedright\endgroup
\end{document}